\documentclass[english,aps,preprint]{revtex4}
\usepackage{verbatim}
\usepackage{textcomp}
\usepackage{color}

\usepackage{babel}

\begin{document}

\title{Peer-review in the Internet age}

\author{Pawel Sobkowicz}

\email{countryofblindfolded@gmail.com}

\begin{abstract}
The importance of peer-review in the scientific process can not be
overestimated. Yet, due to increasing pressures of research and exponentially
growing number of publications the task faced by the referees becomes
ever more difficult. We discuss here a few possible improvements that
would enable more efficient review of the scientific literature, using
the growing Internet connectivity.
In particular, a practical automated model for providing the referees with 
references to papers that might have strong relationship with the work under review, 
based on general network properties of citations is proposed.
\end{abstract}
\maketitle

\section{introduction}

The process of validation of scientific results, especially at the
stage of publication is one of the most important  parts of scientific methodology, 
ensuring quality of the
published work. Current discussions aimed at improving peer-review methods
point out several problems that face the journal editors and referees
(\citet{alberts08-1,raff08-1}). Moreover, the continuously increasing
competitive pressure of securing own job, grants, publications, gives
the task of reviewing others' work an odium of a chore. Both technical
and ethical problems connected with the peer-review process are widely
recognized, and there are publications and web sites aimed at giving
guidance to those who, either as the authors or as the referees,
participate in the process, for example \citet{wager02-1,rockwell05-1}.

One way of improving the process is in helping the editors is selection
of the referees who would have knowledge necessary to assess the 
submitted publications. Such solutions have been discussed already, 
for example by \citet{rodriguez06-1}. These ideas are based on the networked 
nature of the scientific communications. 

In this paper we turn to the other side of the process, namely
providing the already chosen referees with tools aimed at improving the speed
and quality of the review process.
 Exponential growth of scientific literature, expressed both in terms
of the page counts of individual journals and in the growth of the
numbers of the journals themselves makes the task of conscientious
publication review ever more difficult. Following current literature,
even within one's own field(s) of research, is quite difficult. Doing
the same for fields that are somewhat remote is close to impossible.
How then can a reviewer be asked to judge if an article contains enough
new discoveries to warrant publication? How can the editors manage
the review process with the necessary attention to detail and yet
with speed and flexibility required by the growth of data? There are
some initiatives, at both organizational and technical levels, that
can improve the situation.

\section{random citing scientist}

The model of \emph{Random Citing Scientist}, introduced by \citet{simkin03-1,simkin05-1,simkin06-1}
has been quite successful in modelling the popularity of research based
on the number of quotations --- using assumptions that have had nothing
(or almost nothing) to do with actual content, merit, and novelty.
It was sufficient to assume a very simple citing process, in which
when a scientist is writing a manuscript he picks up $m$ random articles,
cites them, and also \textbf{copies} some of their references, each
with probability $p$. It turns out that the statistics of citations
obtained within such model are very close to the actual data. Much
better than in a purely statistical model. For example, there was
--- as in real life --- a large fraction of papers with high number
of citations. In fact, the model has reproduced remarkably well network 
structure of connections between the publications.

The model has broken an unwritten taboo, treating \emph{scientists}
as subjects of research. Worse yet,  \citet{simkin03-3,simkin04-1} have analysed
the occurrences of errors in real citations, such as misprinted age numbers,
or misspelled names. It turned out that quite often there are whole
chains of such wrong citations, showing clearly that the assumption
of \textbf{copying without actually reading} may well be a valid one. One of
the papers was titled: \emph{Read before you cite!}.
If the authors do not read the papers they refer to, what can be expected
from the reviewers?

The initial model has been later expanded to allow for preference to
cite recent articles by \citet{vazquez00-1,vazquez01-1}, 
giving even better fit to observed citation statistics. 
One may ask: what are these models and proofs that scientists
sometimes do cite without reading have to do with peer-review process? 

The answer is that the simple model might teach us how to improve
the process of finding how the reviewed paper stands in comparison
to its {}``competitors'' --- works dealing with similar subjects.
The network structure of links suggested by the model allows some
degree of automation of the process of looking for works that share
the same background with the checked one, and which might
share the research directions and relevant results.

\section{automating the discovery of competitive works}

We  expect that the author(s) should be well versed
in the current state of literature dealing with the subject of their
own research. We also expect the list of references for each article to provide complete 
information necessary to position the work with respect to other developments 
in the field. 
Unfortunately, as the previous section indicates, these assumptions may be invalid. 
In my experience as a referee
I have found quite often that what I thought to be crucial papers
for a given topic were missed by authors of submitted articles. Not
because of ill-will or disrespect towards someone's work, but because
the authors genuinely missed some work, that should be considered
and cited. 
In the age of Internet search engines, databases and
electronic libraries the task of looking up what may be relevant and important is much easier than, 
say, 30 years ago.
On the other hand the sheer number of publications makes it difficult
to find and read all the truly important works. 
When such omission is been pointed out by the referee, in most cases 
the authors agree to extend their list, acknowledging ``valuable suggestion''.   
Is it because agreeing with the referees is seen as the best way to ensure speedy publication?

If missing vital references happens to authors, who know their subject by heart, the situation
must be worse for the reviewers. They are usually chosen from scientists working in 
fields of research close to the paper's subject. 
Not \emph{too close} though, to avoid conflict
of interest. This forces the referees to brush up on the topic of the reviewed paper
to provide real insight.
When we recall that many journals give very
small time windows for the referee answer, the task becomes quite difficult. 
The practical question becomes: how can
we ensure that the reviewer has access to as much as possible of current
literature dealing with the same subject (or very close ones) as the
work he has to check?

In many cases the referee relies not only on his own experiences,
but mainly on the quotations provided by the authors of the reviewed paper. 
This is fine, as the
first step, but not enough. As noted above, the authors might miss, whether accidentally
or on purpose, some papers. 
For example those that show ``the other side of the story''. especially if the subject is controversial.
And  if the referee plans to spend limited time to complete the review, he or she might simply not have the
time to hunt down the missing links. Thus, the evaluation of the innovativeness
and creativity of the reviewed work has to be compromised. For example some results might be treated as novel, even though they have been published elsewhere --- but missed by the referee.

Let's consider how to use the network-based model of Random Citing Scientist
to our advantage. What we propose is that the journal editors do a little preparatory
work before sending the paper to the referee. What kind of work? Simply
take one step \emph{down} the citation network and then one step \emph{up} and in this way 
find the  papers that share a subset
of the same references as the article submitted for review. 
Such search and selection is rather
easy, and in some research fields (for example physics) it can be done with freely
accessible Internet engines. Where no such engines are available, the editors
might use more advanced, fee-based databases. The search should be
directed, reflecting the network structure of citation statistics. An example of such method is given below.

Usually,
any scientific work cites some {}``ground breaking'', classical papers in their
field, or comprehensive reviews --- for the purpose of establishing
the research topic within a more general framework. 
Quite often there's little ``direct'' link between these papers and the results of a publication. 
It is likely, that these ``general'' references are exactly those that the authors copy from other 
papers in the field. The unwritten rules that state ``if you write about such and such topic they 
you should cite this or that review'' are quite natural and human. 
These references therefore add only a little little to the truly novel aspects of the publication under review.

The second class of references consists of much more recent works, cited because of 
their direct relevance
to the research. They would contain sources of data, details of 
experiments or proposed theories used in analysis. Any competitive work would likely 
refer to the same works, to answer similar questions.

How to automatically tell which of the references belongs to each group, without involving human intelligence? 
One of the ways of approximating the split might be via very simplistic criterion of the age of the paper. 
We might assume that the second group would contain
all references not older than 3 years. 
This number is, of course, completely arbitrary, and might be different in physics than in biochemistry. 
But it provides a good starting point in automating the analysis. Moreover, it is
easily incorporated in any search engine. 

After selecting the ``active'' references the next step is to search
for the papers that use the same sources, and to order the results of
the search by the degree of commonality of the whole set of references.
In an  ideal world, two {}``competitive'' papers on the same subject
would have the same set of starting blocks. In practice, this does
not happen, but the results of such search can be quite helpful.
The networked structure reveals often more information than keyword based search.

To test the described method I have done  a  small scale {}``home experiment''.
 I have chosen one of my  own solid state
physics papers \citep{sobkowicz90-1} to try to look into what 
papers would be found by the process described above and if such results would
be valuable to a hypothetical referee. This exercise had all
the benefits of hindsight, and thanks to long time that has passed since publication
it was possible to see ``future'' developments. Technically the search 
was made possible thanks to the SAO/NASA
ADS Physics Abstract Service (\texttt{http://adsabs.harvard.edu}).
By following the procedure described above, I was not only able to
find most of the works of my {}``competitors'' of that time that
I was aware of. I found also one interesting article that I have missed at the
time of writing. The results from the automatic \emph{down-and-up}
search were in fact much better than a Google Scholar search based
on keywords from the paper's title. Of course nothing prevents from
combining the selection tools for better results.
I would encourage all Readers to try this method at least once, either for a paper
that they were asked to review or for one of their own works. Especially if the 
field of research is far from physics --- my lack of knowledge makes it impossible to
evaluate the usefulness of the method for disciplines such as bio-sciences or 
medicine.\footnote{The author would be grateful for information on results of such individual experiments, for example through e-mail.}

Comparative success of the initial experiment suggests that the process might be useful for the process 
of peer review. The list of papers found could be attached automatically by the 
editors in their requests for the referee's opinion. This would speed up the 
review, by making it easier to compare the various approaches to the reviewed topic.
 Perhaps in disciplines other than physics
the scope of the search should be chosen differently, by looking back
into longer periods or by using more stringent culling of  papers found,
for example by enforcing at least some keyword conditions as well. The results
will never be perfect. But on the other hand, I can imagine that a
referee who gets from a journal editor, attached to a request for
a review, a sorted list of possible suggestions of comparative works,
might find it helpful in any discipline. At worst he or she would
ignore the list. At best, it would help him or her to provide a more
accurate evaluation and positioning of the reviewed work within the
current field. %

\section{shared libraries and history of reviews}

The prevailing requirement: \textbf{publish or perish} forces most scientists
to focus on getting their work in print. Moreover, in most countries
 the work of a scientist or institution is evaluated not
by what they publish\footnote{It is clear that the committees and administrators 
who are responsible for
the evaluation can not check the scientific content themselves. They
rely on the very process of peer-review of publications.%
} but \textbf{where} it is published. Placing the article in one of
the high impact journals {}``earns'' in some places much more {}``merit''
than publication in a journal that has  small, local circulation or is  
too limited in scope. The ISI Impact Factor
from Thomson Scientific or Article Influence\texttrademark{} Score(AI)
from \texttt{http://www.eigenfactor.org} measure of a journal's prestige
based on per article citations. Many governing bodies use these to
rank the achievements of individual researchers. Should we wonder
that the high influence journals are flooded with submissions? As
the scientists are fighting, literally, for their (scientific) life,
to expect self-restraint is preposterous.

In such situation the task of the referees becomes even harder. The
rejection level of these high impact journals is very high. What is
the advice given to authors of rejected papers? Most often: re-work and re-submit
--- to a journal with lesser impact. Sometimes this process is repeated
quite a few times. The optimists would say: until the right journal
for the work in question is found. Pessimists: until one finds referees
that are friendly or simply too lax to bother with detailed reviews. 

While the proper matching of articles and journals is a desirable
and valuable goal, we have to remember that the mechanisms to ensure
this goal must take into account the fact that we are dealing with people,
with their individual intentions and emotions, and not just mechanical process of purely
objective evaluation. Thus the processes should be resilient to
most human  weaknesses. Automation and procedural improvements 
shall not be successful  in recognition of great,
ground breaking scientific contributions, but there is a lot that can be improved
in  the iterative process of repeated submissions and reviews.
The most obvious action would be to break down the barriers between separate
journals and publishers. While in some cases the referee pools are
separate, the assumption that for a given field of expertise there
are only so many experts that have the necessary knowledge. This would
allow re-use of the work already done, for example the evaluation done for
\emph{Nature} or \emph{Science}, where the referee may have disqualified
the paper in question for a particular  journal, but suggested other venue.
Not only the number of iterations could be lessened, but the changes
requested at each stage would be more coherent, ideally leading to
an article of improved quality finding the right journal faster. Such
solution is already in place in some journals which are owned by the
same publisher, as mentioned by \citet{alberts08-1}. 

The second issue is the reviewer anonymity. While there are arguments
for such anonymity, there are also reasons to shift to non-anonymous
reviews. What is lost in the openness and candour of review due to
lack of anonymity, may be balanced by increase of responsibility that
comes when you sign any document. Additionally we may think about
using the flexible nature in Internet publications to attach (anonymous
or not) peer reviews and authors' responses alongside research articles.
This is already done by some journals, for example by those from BioMed
Central. While quite difficult in the {}``paper age'', the documented
flow of comments, corrections and improvements is quite easy to
present in electronic form. Moreover, it could become the seed for
discussion forums related to the published work. 
The review and
ranking done on-line can bring huge advances. A good example is \emph{Faculty
of 1000 Biology and Medicine} where over 4500 leading researchers
and clinicians share their expert opinions by highlighting and evaluating
the most important articles in biology and medicine (\texttt{http://www.facultyof1000.com/}).
All these ideas use of
technologies available today and would improve the quality of published scientific
work.

\section{conclusions}

None of the solutions for smooth-lining the peer-review described
in this paper is revolutionary. For all of them the technical basis
already exists, maybe in limited form, but certainly applicable. The
ideas of pooling reviews for a group of journals and of sharing the
results as well as co-publishing the reviews and original papers are
already present. The key to improvement lies in the widespread usage
of the tools for sharing information that are available thanks to
today's technologies. The secondary outcome of this would be the improvement
in the workflow of the referees, which would enable more people to
participate meaningfully in the process without undue effort. Both
younger and senior researchers would benefit from such tools and methods.
The goal would be to give the referees all the help to allow them
to concentrate on in-depth evaluation of the papers in question. This
would allow, hopefully, to keep the scientific standards high, despite
the exponential growth in volume of our {}``production''. 

\end{document}